# Generation of Coherent Quantum Light from a Single Impurity-Bound Exciton

Yuxi Jiang,[1,2,*] Christine Falter,[3,5] Robert M. Pettit,[1] Nils von den Driesch,[4,5] Yurii Kutovyi,[4,5] Jasvith Raj Basani,[1,2] Amirehsan Alizadehherfati,[1,2] Alexander Pawlis,[4,5] and Edo Waks[1,2,†]

[1]Institute for Research in Electronics and Applied Physics and Joint Quantum Institute, University of Maryland, College Park, Maryland 20742, USA

[2]Department of Electrical and Computer Engineering, University of Maryland, College Park, Maryland 20740, USA

[3]Peter-Grünberg-Institute (PGI-9), Forschungszentrum Jülich GmbH, 52428 Jülich, Germany

[4]Peter-Grünberg-Institute (PGI-10), Forschungszentrum Jülich GmbH, 52428 Jülich, Germany

[5]JARA-Fundamentals of Future Information Technology, Forschungszentrum Jülich and RWTH Aachen University, 52062 Aachen, Germany

Corresponding author: Yuxi Jiang, Edo Waks

[*]yxjiang@umd.edu

[†]edowaks@umd.edu

## Abstract

Impurity-bound excitons in II-VI semiconductors are promising optically active solid-state spin qubits that combine exceptional optical quantum efficiency with a low noise spin environment. Previous studies on single impurities in these materials relied on incoherent optical excitation to generate photons. However, many quantum applications require resonant driving of quantum emitters to precisely control optical transitions and maintain coherence of the emission. Here, we demonstrate coherent optical emission of quantum light from a resonantly driven single impurity-bound exciton in ZnSe. The resonantly driven emitter exhibits bright quantum emission that preserves the phase of the resonant drive. It also features an intensity-dependent nonlinear phase shift at low photon numbers. Resonant excitation enables direct measurement of the Debye-Waller factor, determined to be 0.94, which indicates high efficiency emission to the zero-phonon line. Time-resolved resonance fluorescence measurements reveal a fast optically driven ionization process, along with a slower spontaneous ionization process having a lifetime of 21 $\mu$s due to charge tunneling from the impurity. These results pave the way for coherent generation of quantum light and low-photon number nonlinear optics through resonant excitation of impurity-bound excitons.

## Introduction

Impurity-bound excitons in II-VI semiconductors are promising candidates for optically active spin qubits due to their bright emission and potentially long spin coherence time[1–11]. Among these materials, shallow donor-bound excitons in ZnSe stand out for their ability to generate indistinguishable single photons[1–3]. Zinc and Selenium also possess a high natural abundance of zero nuclear spin isotopes that can be isotopically purified to create an almost entirely nuclear spin-free environment, which is ideal for electron spin qubits[4]. Moreover, ZnSe can be grown in epitaxial thin films, that can be patterned into nanophotonic devices to enhance light-matter interactions and single-photon emission efficiency[3,5,6]. Furthermore, these quantum emissions can be converted to telecom wavelengths for fiber-based quantum networks through quantum frequency conversion[12,13].

Virtually all optical studies of impurity-bound excitons in ZnSe have relied on incoherent excitation of the emitters using above-band pumping[2,3,5–7]. This approach suffers from random time-jitter associated with carrier relaxation and induces fluctuating local electric fields, both of which create optical decoherence and broaden the emission linewidth. Furthermore, incoherent excitation cannot probe the internal energy levels of the impurity-bound excitons. Resonant excitation mitigates these problems by selectively driving transitions between the quantum states[14–18]. In this approach, a narrow-bandwidth tunable laser pumps a bound exciton on resonance, generating resonance fluorescence. Detecting this coherent emission is challenging because it is masked by scattering and reflection from the resonant laser. Moreover, resonant excitation can alter the emitter charge state, quenching its fluorescence and making it even more difficult to observe resonant emission[2,10].

In this letter, we demonstrate coherent quantum emission from a single impurity-bound exciton under resonant excitation. The impurity is a Chlorine atom embedded in a ZnSe quantum well, which is integrated into a nanopillar device exhibiting bright single-photon emission[2]. We resonantly drive the impurity and observe bright quantum light emission, verified through second-order correlation measurements. Using polarization interferometry, we also validate that the emission retains phase coherence with the incident pump, an important property for entanglement distribution[19] and squeezed light generation[20]. Furthermore, we demonstrate that the reflected field exhibits a highly intensity-dependent interference at low incident optical powers, with potential applications in low-photon-number nonlinear phase gates[20–22]. Through resonant excitation, we also measure the Debye-Waller factor of the impurity-bound exciton emitter to be 0.94, one of the highest values reported for defect emitters. Time-resolved measurements of resonantly pumped bound exciton fluorescence reveal a rapid, optically induced ionization process, and a slower spontaneous discharging of the impurity state. A low-power above-band laser can stabilize the impurity charge state and recovers the resonance fluorescence emission on a rapid timescale of 9.3 ns, enabling fast optical gating[23] of the coherent quantum light source. Our results provide a new optical toolbox for generating coherent emission and directly manipulating the electronic states of impurity-bound excitons.

## Main text

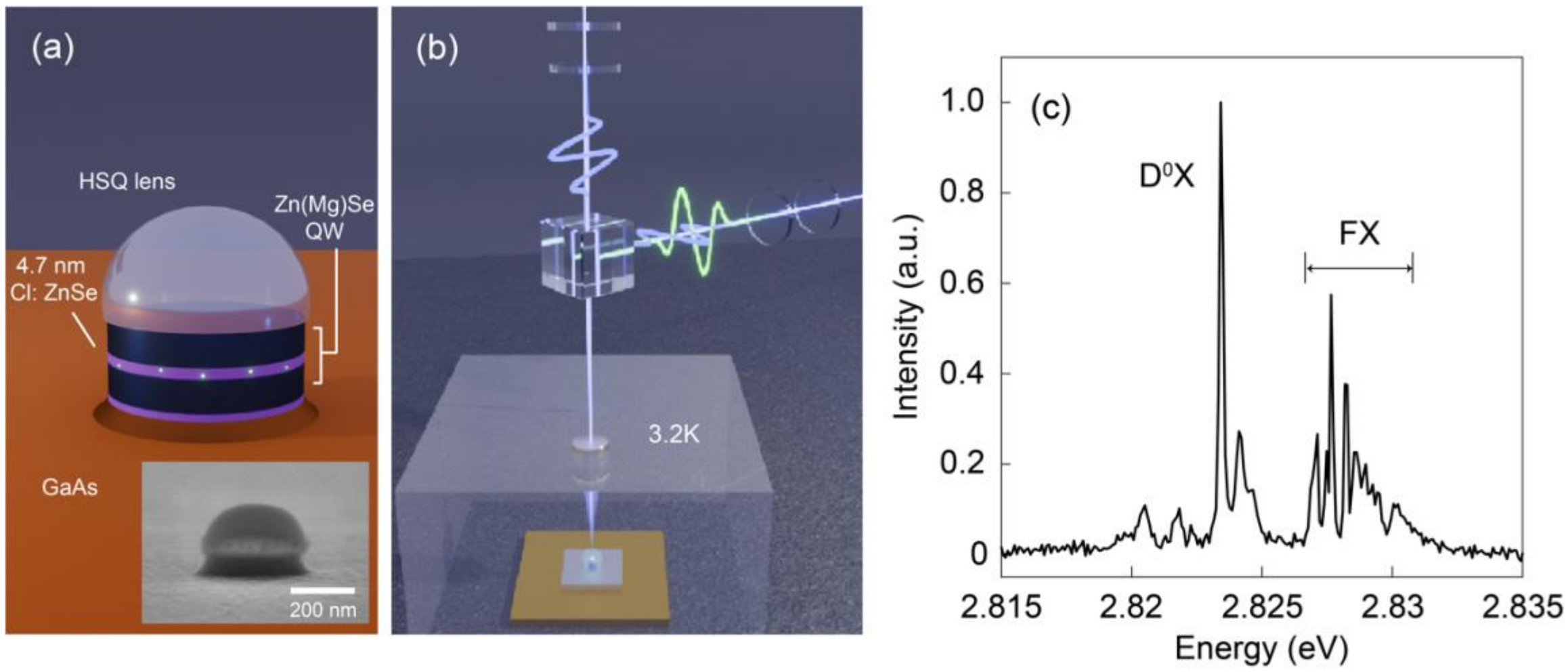

**Figure 1. Experimental setup and optical measurements of an impurity-bound exciton.** (a) A schematic of the nanopillar device consisting of ZnMgSe barrier layers enclosing a 4.7-nm-thick quantum well (QW) of Cl-doped ZnSe on a GaAs substrate. The nanopillar is topped by a hydrogen silsesquioxane (HSQ) hemispherical nanolens. The inset shows a scanning electron microscopy image of the fabricated device. (b) The resonant excitation and fluorescence detection scheme. The laser excites the sample within the cryostat through a confocal microscope setup. Two sets of waveplates can independently control the polarization of the excitation laser and the output light reflected from the sample. (c) The photoluminescence spectrum of the ZnSe nanopillar under above-band excitation at 3.06 eV (405 nm), featuring emission lines corresponding to free-exciton (FX) and the donor-bound exciton ($D^0X$).

## *Experimental setup and photoluminescence characterization*

Figure 1a shows an illustration of the device used in this work. The initial wafer is composed of a ZnMgSe/ZnSe/ZnMgSe quantum well structure on a GaAs substrate, with a Cl delta-doped layer in the ZnSe quantum well. Since atomic radius of Cl and Se are similar (100 pm and 120 pm), Cl is likely incorporated as a substitutional donor on the Se lattice sites. From this wafer, we fabricate a nanopillar structure with a diameter of 250 nm and a height of 75 nm, with an HSQ lens on top to enhance the out-coupling efficiency of the fluorescence emission[3] (see Methods for detailed descriptions of the optical device). The inset of Figure 1a shows a scanning electron microscopy image of the fabricated device.

Figure 1b shows the optical measurement setup used to assess the device emission. We cool the device to 3.2 K using a closed-cycle cryostat and excite it through a confocal microscope. We resonantly excite the sample using a tunable laser whose polarization is controlled by a half-wave plate and a quarter-wave plate. We detect the resonance fluorescence in a cross-polarized detection scheme, which rejects stray reflections of the resonant laser from the sample surface. The Methods section contains a detailed description of the measurement technique.

We initially characterize the device through photoluminescence. We pump the device using a 3.06 eV (405 nm) laser, whose energy is higher than the ZnSe and ZnMgSe bandgaps. Figure 1c shows the resulting spectrum. We observe a group of peaks ranging from 2.827 eV to 2.832 eV that correspond to the free-exciton emission (FX). A discrete narrow peak at 2.8233 eV labeled as $D^0X$ corresponds to a single donor-bound exciton emission[3]. Supplementary Material Section I and Figure S1 provides additional photoluminescence measurements that characterize the emission polarization and determine the above-band saturation power to be 2.9 $\mu$W.

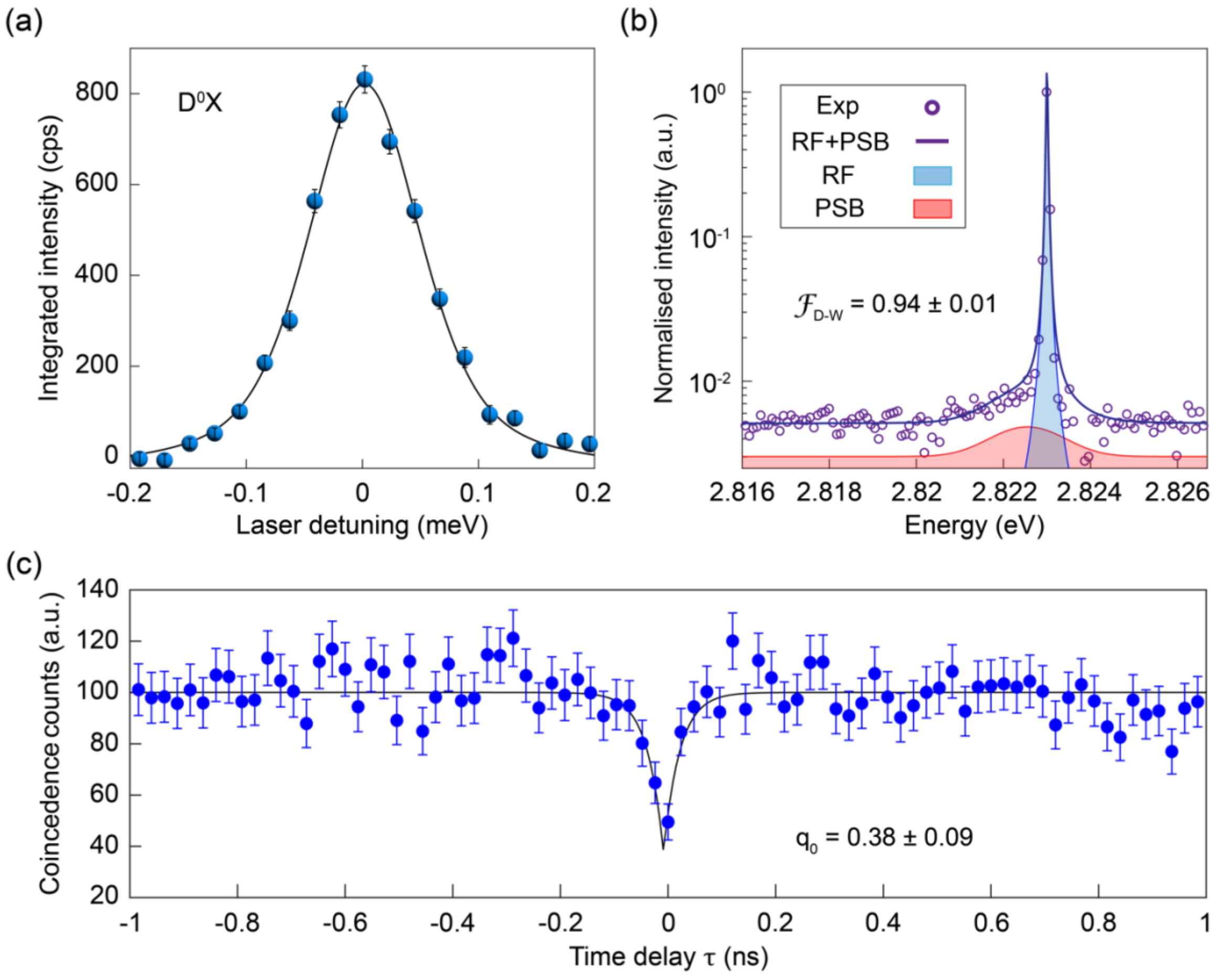

**Figure 2. Resonance fluorescence from an impurity-bound exciton in ZnSe.** (a) Resonance fluorescence intensity as a function of laser detuning from the bound exciton resonance. Solid markers indicate measured data, while the line represents a numerical fit to the Voigt function. Error bars represents the statistical uncertainty and propagating errors in detected events. (b) Fluorescence spectrum of the resonantly pumped impurity-bound exciton. The open circles represent the measured data, and the solid line shows the theoretical fit to a sum of a Lorentzian function for the resonance fluorescence and a Gaussian function for the phonon sideband. The blue and red shaded areas are a plot of the Lorentzian and Gaussian functions from the fit individually. RF: Resonance Fluorescence, PSB: Phonon sideband. (c) Second-order correlation measurement of the resonance fluorescence.

### *Resonance fluorescence from a single Cl-impurity-bound exciton*

We next use a tunable laser to resonantly excite the bound exciton line and measure the emission signal through spectrometer. To suppress background light from the direct reflection of the pump laser, we collect the output light from the sample in a cross-polarized state with respect to the input light (see Methods). This cross-polarization filter has an extinction ratio exceeding $10^6$. The laser has a stabilized power of 10 nW, measured before the objective lens, to maintain a weak excitation condition of the bound exciton.

When we only use the resonant laser to excite the bound exciton, we observe little emission from the state. This observation is consistent with previous work showing that a weak above-band pump can stabilize the quantum well charge environment[2]. Therefore, we inject an above-band (3.06 eV) laser to excite the sample, while resonantly pumping the bound exciton state. We discuss the needed power of this weak above-band laser in the Supplementary Material Section II and Figure S2. We determine that an above-band laser with optical powers in a range from 3 nW to 32 nW (0.1% to 1.1% of the above-band saturation power) can efficiently stabilize the impurity state while producing negligible photoluminescence from the state.

Figure 2a shows the integrated intensity of the resonant emission from the bound exciton as a function of laser detuning (see Supplementary Material Section III for details of extracting the emission intensities from the measured spectra). The spectrum shows a clear resonant behavior where emission is maximum at the bound exciton energy, and quickly falls off with detuning. In contrast, we perform the same scan with no above-band laser where the impurity state is discharged and optically inactive, and the result only shows background light due to imperfect cross-polarization (see Figure S3 in the Supplementary Material Section III). The solid line in Figure 2a is a numerical fit to a Voigt function, defined as a convolution between a Lorentzian and Gaussian distribution. This function represents a good description of the emitter linewidth when there is inhomogeneous broadening[24,25]. The resonance fluorescence spectral resolution is significantly finer than the spectrometer resolution limit (0.09 meV), as detailed in the Supplementary Material Section III. From the fit of the resonance fluorescence peak, we obtain a linewidth of $(133 \pm 11.6)$ $\mu$eV, which is 28-times broader than the lifetime limited linewidth of 4.7 $\mu$eV (calculated from measured radiative lifetime of $\tau_{\mathrm{rad}} = 141 \pm 1$ ps). The pumping power of 10 nW is well below the resonant saturation power (~ 0.1 $P_{sat}$), as determined by the power-dependent measurement in the Supplementary Material Section III. Therefore, power broadening is unlikely to be the main cause of the observed broadened linewidth. We attribute the broadened linewidth to spectral wandering[26], which may arise from charge fluctuations of defects and trap states near the Cl impurity[27].

Figure 2b shows the resonant excitation spectrum when the laser is on resonance with the bound exciton state, plotted on a semilogarithmic scale. The open circles represent the measured spectrum which exhibits a bright emission peak at the bound exciton resonant pumping energy of 2.823 eV, with spectrometer resolution limited linewidth. This emission peak represents the resonance fluorescence emission from the bound exciton. We also observe a broader peak detuned from the central emission line, which we attribute to the phonon sideband. This phonon sideband, which is the only one observable under this excitation condition, likely arises from the acoustic phonons in ZnSe, whose coupling strength to the emitter is enhanced by quantum well confinement[28–30], while the longitudinal optical-phonon coupling is suppressed by the confinement[31].

We fit the measured spectrum to a sum of a Lorentzian function, representing the resonance fluorescence, and a Gaussian function, which serves as a good model for the phonon sideband[32]. The blue and red curves in Figure 2b show the fitted results from the Lorentzian and Gaussian functions, respectively, while the purple solid line shows the total fit, which exhibits excellent agreement with the measured data. From the fit, we can extract the Debye-Waller factor of the emitter, which is the ratio of the elastically scattered fluorescence intensity to the total emission intensity of the emitter. We calculate this factor as $\mathcal{F}_{\mathrm{DW}} = I_{\mathrm{RF}}/(I_{\mathrm{RF}} + I_{\mathrm{PSB}})$, where $I_{\mathrm{RF}}$ and $I_{\mathrm{PSB}}$ are the areas under the fit for the resonance fluorescence and phonon sideband, respectively. Using the data from the fit in Figure 2b, we obtain $\mathcal{F}_{\mathrm{DW}} = 0.94$, which is among the highest values for reported defect quantum emitters[33–35]. Supplementary Materials Section III and Figure S4 provides additional temperature-dependent measurements of the resonantly pumped bound exciton, showing that the sideband remains unchanged in intensity up to approximately 10 K and then exponentially increases. Thus, our measurements at 3.2 K are at a sufficiently low temperature to minimize the phonon sideband and accurately estimate the Debye-Waller factor of the bound exciton.

We validate the quantum nature of the resonance fluorescence by performing the second-order correlation measurement. We excite the donor-bound exciton state on-resonance with a laser power of 20 nW. The signal is detected using an intensity correlation setup with two superconducting nanowire single photon detectors (see Methods for details).

Figure 2c shows the measured second-order correlation. At zero time-delay, the spectrum exhibits clear anti-bunching, demonstrating the quantum nature of the emission. We fit the measurements to an exponential curve of $A\left(1-(1-q_0)\mathrm{e}^{-|\tau|/\tau_0}\right)$, where $\tau_0$ is a parameter determined by the radiative decay rate and the pumping power[36]. The parameter $q_0$ represents the zero-time delay correlation value. From the fit, we obtain $q_0 = 0.38 \pm 0.09$, which falls below the threshold of 0.5, validating that this is a single emitter. The single photon purity is limited by the background noise, which originates both from the stray reflection of the resonant laser and some free-exciton and trion emission that overlaps with the bound exciton emission due to above-band laser stabilization. We define the signal to noise ratio as $R = S/(S+B)$, where $B$ is the background count rate and $S = T - B$ where $T$ is the total count rate. We measure these values (See Supplementary Material Section IV), and attain $R = 0.85$. We use this value to correct for false coincidences due to backgrounds and attain the fundamental single photon purity of the emitter (see Supplementary Material Section IV). Using this correction method, we obtain a background-corrected $q_0$ of 0.13.

## *Coherent emission and polarization interferometry*

We next investigate the coherence of the resonance fluorescence emission. Under resonant excitation, the resonance fluorescence field is given by $E_{\mathrm{RF}} = \frac{\gamma}{-\mathrm{i}\Delta+\frac{\gamma}{2}}\sqrt{\eta}E_0$,[37] where $\gamma$ is the atomic transition decay rate, $\Delta$ is the laser detuning, $E_0$ is the electric field amplitude of the excitation laser, and $\eta$ is the collection efficiency of the fluorescence. We derive this equation in the Supplementary Materials Section V. The equation shows that the complex amplitude of the resonance fluorescence field varies as a function of the laser detuning. Specifically, the phase of the resonance fluorescence at a detuning of $\Delta \ll -\gamma$ experiences a π-phase shift relative to that of $\Delta \gg \gamma$. As the excitation energy is swept across the resonance of the zero-phonon line of the bound exciton, there is a rapid phase transition between these two conditions. We can measure this phase by mixing the fluorescence with a local oscillator, resulting in either constructive or destructive interference depending on the detuning to the bound exciton.

To measure this phase, we perform single-beam polarization interferometry[21,22]. We mix the resonance fluorescence of the emitter with a local oscillator generated by taking a fraction of light from the excitation laser. This mixing is achieved by rotating the input quarter-wave plate in the cross-polarization setup, which modulates both the intensity and relative phase of the local oscillator (see Methods for the experimental implementation).

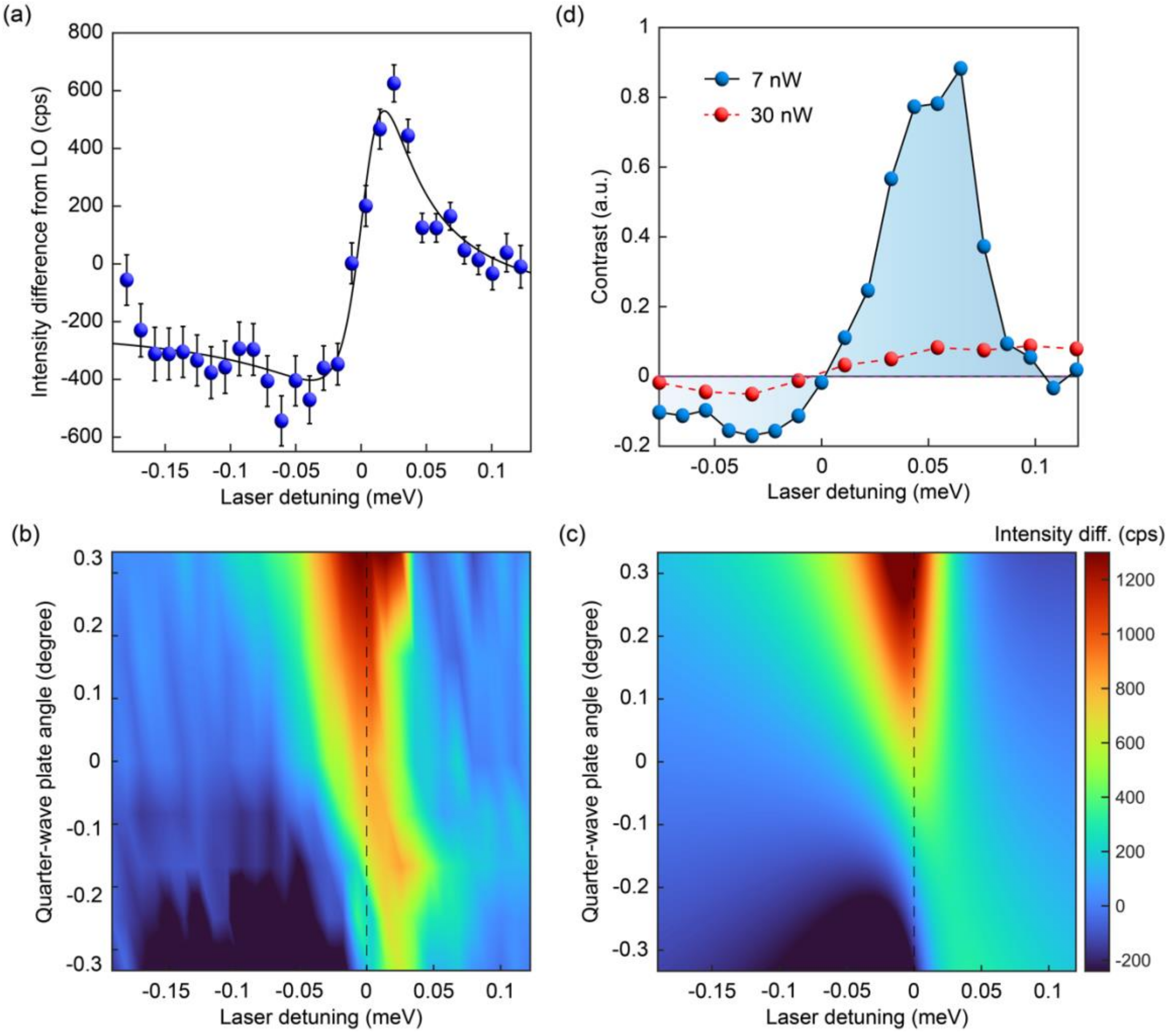

**Figure 3. Polarization interferometry demonstrating the phase coherence of the resonance fluorescence.** (a) Measured reflection intensity difference (denoted by the blue solid dots) between the total emission and the local oscillator plots as a function of laser detuning, showing a Fano-shape interference pattern when scanning across the bound exciton resonance. In this experiment, the quarter-wave plate in the polarization setup is rotated $-0.3$ degree from the optimized cross-polarization condition. The line shows a numerical fit to the theoretical model of this interference effect. Error bars account for the statistical uncertainty in detected events and propagating errors in the subtraction. (b) Measured and (c) calculated reflection signal intensity difference as a function of laser detuning at different quarter-wave plate angles. The dashed line denotes the zero detuning energy. Color bar denotes the intensity

difference from the local oscillator. (d) Resonant scanning spectrum at different resonant laser powers of 7 nW (blue dots and the solid line) and 30 nW (red dots and the dashed line).

Figure 3a shows the results of the polarization interferometry measurement at a quarter-wave plate angle of $-0.3$ degree. We measure the resonant laser reflection intensity $I_{\mathrm{tot}}$ and subtract it from the local oscillator intensity $I_{\mathrm{LO}}$, obtained using only the resonant laser without above-band excitation. In the figure, the dots represent the measured intensity difference. The resulting spectrum exhibits a Fano-resonance pattern[38]. At higher energies, we observe a peak with the maximum energy position shifting from the original resonance fluorescence peak position (centered at the zero-energy detuning), indicating constructive interference. At lower energies, we observe destructive interference, where the $I_{\mathrm{tot}}$ from the mixing of the fluorescence with the local oscillator is less than the $I_{\mathrm{LO}}$. This rapid transition from constructive to destructive interference validates the phase coherence of the emission. The solid line shows the curve fit to the measured data from a theoretical model, which exhibits good agreement and reveals the $\pi$-phase shift of the reflected light across zero-energy detuning, as detailed in Supplementary Material Section V.

Figure 3b-c shows the measured and theoretically calculated interference spectrum as a function of the waveplate angle and the detuning energy (see Supplementary Materials Section V for a detailed description of the theoretical model). The experimental data and theoretical model exhibit excellent agreement in their interference spectrum. At angles below $-0.1$ degree, we observe a clear interference transition from lower to higher energies. When rotating the waveplate to positive angles, a constructive interference effect appears at red-detuned energies, while only weak destructive interference occurs at blue-detuned energies. The asymmetricity originates from the interference between the real part of the resonance fluorescence and the local oscillator, as described in more detail in the Supplementary Material Section V and Figure S9.

We next measure the intensity dependence of the interference pattern. We define the interference contrast as $C = (I_{\mathrm{tot}} - I_{\mathrm{LO}})/I_{\mathrm{LO}}$, where $I_{\mathrm{tot}}$ is the total reflection intensity and $I_{\mathrm{LO}}$ is the local oscillator intensity. Figure 3d plots this contrast as a function of detuning for a low power of 7 nW and a high power of 30 nW (here we set the quarter waveplate angle to $-0.5$ degrees). The reflection signal exhibits strong nonlinear behavior, showing high-contrast interference at the low power that diminishes at the higher power. This marked nonlinear effect results from the saturation and the quenching of the bound exciton emitter at higher powers[17,25]. A further discussion of the nonlinear excitation power is provided in the discussion section.

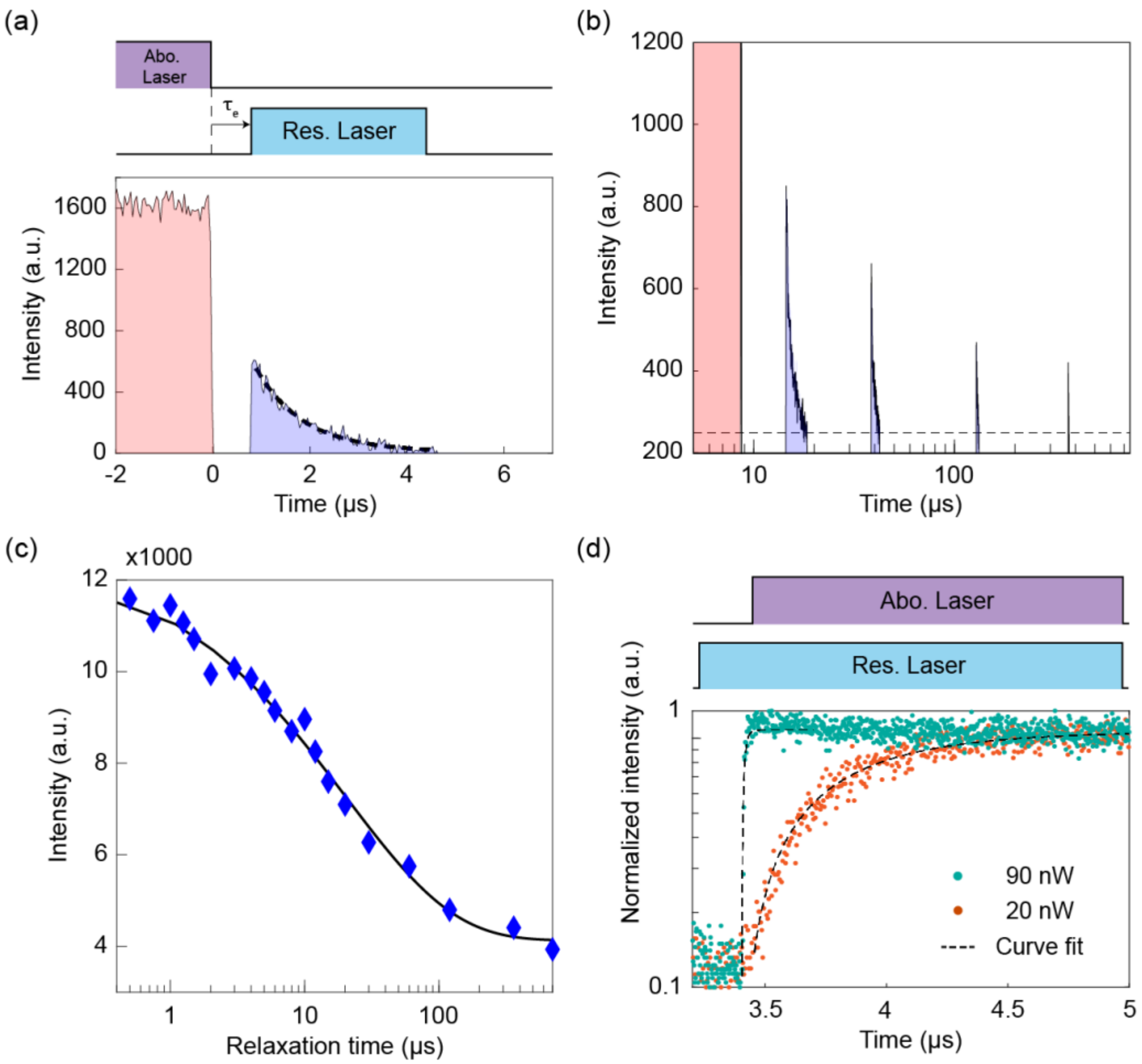

**Figure 4. Dynamics of the impurity-bound exciton resonance fluorescence.** (a) Upper panel: Pulse sequence for time-resolved resonance fluorescence. Lower panel: the corresponding fluorescence intensity as a function of time under two lasers excitation. The red-shaded area indicates photoluminescence from above-band excitation, and the blue-shaded area indicates resonance fluorescence from the bound exciton under resonant pumping. There is a 1.2 $\mu$s delay between the resonant laser pulse and the above-band excitation. (b) Time-resolved resonance fluorescence for four different delays between the above-band and resonant laser. (c) The integrated intensity of the resonance fluorescence as a function of the delay time. Blue markers show the measured data, and the solid line shows a stretched exponential decay fit to the measurement. (d) Upper panel: the laser modulation pulse sequence. Lower panel: Time-resolved fluorescence intensity as a function of time under the above pumping scheme at two different above-band laser powers (90 nW, green dots, 20 nW, red dots). Res., Resonant; Abo., Above-band.

*Dynamics of the impurity-bound exciton resonance fluorescence*

To better understand the dynamical response of the resonance fluorescence signal, we drive the emitter using laser pulses, with the pulse sequence illustrated in the upper panel of Figure 4a. We first pump the emitter with a 5 $\mu$s above-band laser pulse with an average power of 40 nW to charge-stabilize the impurity. We then probe the impurity by exciting it with a 4 $\mu$s resonant laser pulse with an adjustable delay time $\tau_e$ to generate the resonance fluorescence. Additional details are provided in Supplementary Materials Section VI.

Figure 4a shows the fluorescence emission intensity as a function of time for a delay of $\tau_e = 1.2$ $\mu$s between the pump and probe pulses. We observe direct photoluminescence during the above-band pumping period (denoted by the red shaded area). We then resonantly drive the bound exciton and measure the resonance fluorescence signal. As shown by the blue shaded area in Figure 4a, the resonance fluorescence initially turns on and quickly decays to zero. By fitting the data to an exponential decay function, we obtain a decay rate of 1.013 $\mu s^{-1}$. This rate is much smaller than the radiative decay rate of the bound exciton state and increases linearly as a function of the resonant power, as shown by the power-dependent measurements in Supplementary Material Section VI and Figure S11. We further discuss this non-radiative decay of the coherent emission under the resonant excitation in the Discussion Section.

In Figure 4b, we show results of the same measurement but for various delay times $\tau_e$ ranging from 0.5 $\mu$s to 200 $\mu$s. The measurements exhibit a decrease in the resonance fluorescence peak intensity as the delay time increases. Figure 4c plots the emission intensity as a function of the delay time. The data is best fit by a stretched exponential decay function with a time constant of 21 $\mu$s. We attribute this decay to spontaneous discharging of the impurity, which may originate from tunneling to nearby trapping or surface states[39]. This spontaneous discharging of the impurity is consistent with the measurements shown in Figure 2a, which demonstrate that the impurity transits to an optically active state using a small amount of above-band light to stabilize the charge.

To determine the time scale of the charge recovery, we utilize the pulse sequence in the upper panel of Figure 4d. Here we first turn on the resonant laser and then turn on the above-band laser. The initial resonant laser quenches the emitter, while the non-resonant laser pulse recharge it, leading to a recovery of the resonance fluorescence signal. Figure 4d shows the resonance fluorescence intensity using this modified pulse sequence for above-band powers of 20 nW and 90 nW. The emission rapidly recovers at the onset of the above-band laser, eventually reaching its maximum value. To obtain the recovery time constants, we fit the two curves to a function of $I(t) = I_0\left(1 - \mathrm{e}^{(-t/\tau)}\right)$, which reflects the population change of the impurity state. The higher above-band power leads to a faster signal recovery, with the time constant decreasing from 200 ns to 9.3 ns for the 20 nW and 90 nW pulses, respectively. These results demonstrate the potential of short above-band pulses to restore charge to an ionized emitter on nanosecond timescales, enabling fast optical switching[17,23] of the coherent emission.

## Discussion

In this work, we demonstrate coherent quantum emission of light from a single Cl-impurity in ZnSe by resonantly driving the bound exciton state. Resonant drive enables us to extract a high Debye-Waller factor of 0.94, among the highest values in solid-state defects. We demonstrate phase interference between this coherent emission and a local oscillator through polarization interferometry. Furthermore, in the time-resolved measurements, the coherent emission dynamics reveal impurity discharging, mitigated by a small amount of above-band excitation.

We now address how much optical power is required to elicit a nonlinear response in the scanning spectra shown in Figure 3d. To achieve single photon nonlinearity, the quantum emitter's extinction cross-section should approach the spot size of the excitation laser. The extinction cross-section is given by $\sigma_{ext} = \frac{3(\lambda/n)^2}{2\pi}\frac{\gamma_0}{\gamma_h}$,[21,40] where $n$ is the refractive index of ZnSe, $\gamma_0$ is the lifetime-transform-limited linewidth, and $\gamma_h$ is the homogeneous linewidth of the bound exciton. Due to the spectral wandering, it is difficult to obtain an accurate homogeneous linewidth for the emitter. Assuming the homogeneous linewidth of the emitter approaches its lifetime-limited linewidth, we calculate an ideal extinction cross-section $\sigma_{ext} = 1.14 \times 10^{-2}\ \mu m^2$. Inhomogeneous broadening should not affect the emitter's extinction cross-section. When the emitter spectrally wanders off-resonance, it does not contribute to the detected signal. Using the inhomogeneous linewidth $\gamma_{inh}$ of the emitter, we instead calculate a time-averaged extinction cross-section, as $\sigma_{avg} = \frac{3(\lambda/n)^2}{2\pi}\frac{\gamma_0}{\gamma_{inh}}$. Using $\gamma_0/\gamma_{inh} = 1/28$ from Figure 2a, we calculate $\sigma_{avg} = 4.0 \times 10^{-4}\ \mu m^2$. This is much smaller than the cross-section $\sigma_{ext}$ and using it would underestimate the emitter's extinction cross-section. The actual excitation spot size can be difficult to estimate from Gaussian beam optics, because the lens at the top of the nanopillar can strongly modify the focus spot. We therefore perform finite-difference time-domain (FDTD) simulations to obtain the electric field distribution near the emitter (See Supplementary Material Section VII and Figure S13 for details), from where we extract a mode area $A_{mod} = 4.2 \times 10^{-2}\ \mu m^2$ of the input light. Comparing this mode area to the extinction cross-section we obtain $A_{mod}/\sigma_{ext} = 3.7$, indicating that the device works in the few photon number regime, but not at the single-photon level. Additionally, spatial misalignment between the bound exciton and the local optical field as well as the pure dephasing of the emitter may further increase this ratio. Integrating the emitter with a cavity can reduce the ratio by enhancing the radiative decay rate through the Purcell effect[5]. These methods will push the device to the actual single photon nonlinear regime.

We next consider the physical processes that could lead to the extinction of the coherent emission shown in Figure 4a. In Supplementary Materials Section VI, we show that the decay rate increases linearly as a function of resonant pumping power. In the weak excitation regime, where the pumping power $P \ll P_{sat}$, the excited state population $\rho_{ee}$ scales nearly linearly with $P$. The measured emission decay (with a rate $\Gamma_{OID}$) arising from a discharging process in the excited state is proportional to $\rho_{ee}$ and to the resonant power ($\Gamma_{OID} \propto \rho_{ee} \propto P$). Two-photon processes, which usually leads to a quadratic dependence of the pump power[41], are therefore not likely to be the cause of this decay. Instead, the emitter likely undergoes an optically induced discharging while resonantly pumping. When we attempt to re-excite the bound exciton resonantly after the decay happens, the emitter remains dark for longer than 20 $\mu$s (See Figure S12 in Supplementary Material Section VI). This measurement demonstrates that the impurity does not recover to the optically active state once discharges, suggesting that the dominant quenching mechanism is emitter discharging. This discharge can be due to several physical processes, including Auger recombination[42], and photo-induced ionization of the neutral donor defect[17,43,44].

We compare the emission properties of the impurity-bound exciton in ZnSe to those of other solid-state quantum emitters. The rapid radiative decay and the high Debye-Waller factor are standout features for these bound exciton emitters. The radiative lifetime of the bound exciton is roughly an order of magnitude shorter than that of typical diamond color centers[45] and single molecules[46], and is comparable to that of strongly confined quantum dots. The Debye-Waller factor of the bound exciton surpasses that of nitrogen-vacancy (NV) centers in diamond (0.04) and other group-IV defects such as the SiV and SnV centers (0.6-0.8)[47,48]. Although the emitter exhibits an inhomogeneously broadened linewidth, this broadening can be further mitigated by optimizing fabrication processes to reduce surface trap states, such as implementing *in-situ* structuring of epitaxial thin films[49]. Previous research has demonstrated quantum interference between photons emitted from two bound excitons in ZnSe, suggesting that the homogenous linewidth of the emitter can be lifetime limited[1]. As atomic defect centers, these bound excitons offer efficient quantum emission combined with high zero-phonon emission efficiency, which was previously reserved for high-quality quantum dots and single molecules.

Looking forward, the Cl-impurity shallow-donor defects not only emit bright coherent emission but also naturally possess a spin ½ ground state, which can act as an optically active spin qubit. The resonant excitation methods we present here provide a direct pathway towards optical control and readout of the electron spin[50,51]. The integration of these emitters with optical cavities could further enable coherent spin-photon interaction[5]. For example, a high quality-factor optical cavity will enable the design of spin-photon hardware, such as a single-photon switch[52] based on a single impurity donor-bound exciton. Additionally, these solid-state quantum emitters are compatible with hybrid quantum photonic integration. The emission wavelength of the bound exciton aligns with that of two-dimensional hBN color centers[53] and is close to the transition wavelengths of trapped-ion systems[54]. Integrated systems will benefit from the bright quantum emission of the bound exciton. Furthermore, these quantum emissions can be efficiently converted to telecom photons through two-stage quantum frequency conversion[12,13], thereby enabling long-distance entanglement distribution for scalable quantum networks. Ultimately, this work opens the possibility for direct optical control of impurity-bound excitons to achieve efficient quantum light sources and spin-light interfaces.

## Methods

*Device description*

The ZnMgSe/ZnSe/ZnMgSe QW structure is grown using molecular beam epitaxy on a GaAs substrate. The 4.7-nm-thick ZnSe QW is enclosed between two 29 nm ZnMgSe barriers, all on top of a 12-nm-thick ZnSe buffer layer between the ZnMgSe barriers and the substrate as shown by Figure 1a. The Cl donors are incorporated in the ZnSe QW by delta-doping with a low sheet density in the order of $10^{10}$ $cm^{-2}$. An abrupt interface between Cl-doped and undoped regions in ZnSe was previously demonstrated in the growth process, indicating negligible Cl-diffusion[55]. To form single isolated Cl-impurities, nanopillars are fabricated in a top-down nanofabrication approach. After sample growth, the nanopillar and nanolens patterns are defined using HSQ negative electron beam resist and greyscale electron-beam lithography. Then the nanopillar structures are fabricated by dry etching via inductively coupled plasma reactive ion etching with a combination of $H_2$/Ar/$CHF_3$ gases, which also forms the spherical nanolens on top of each pillar. Finally, the sidewalls exposed during the dry etching step are polished with a potassium dichromate solution and afterwards passivated with a 10 nm thick AlOx layer grown by atomic layer deposition (ALD).

We determined the out-coupling efficiency of the nanopillar devices using FDTD optical simulations (Lumerical). The emitter was modeled as an electric dipole source positioned at the center of a Zn(Mg)Se nanopillar. We calculated the fraction of emission collected by an objective lens with a numerical aperture (NA) of 0.68. The simulations indicate an out-coupling efficiency of 0.16 for the nanopillar with an HSQ lens, decreasing to 0.12 in the absence of the lens. For comparison, an emitter in the unstructured quantum well membrane exhibits a significantly lower collection efficiency of 0.02.

*Optical measurement setup*

Figure 1b shows the schematic of the experimental setup. We mount the sample in a closed-loop cryostat (Attocube, attoDRY 1000). A free-space confocal microscope is used to both excite the sample and collect the reflected optical signal, with an objective lens of NA = 0.7. The above-band excitation is achieved by a laser diode (Thorlabs, LP405-SF10). The resonant excitation and the laser scanning are performed using a tunable diode laser (TOPTICA, DL pro). This tunable laser is sent through a polarization-maintained single-mode fiber and transmits through a linear polarizer, a half-wave plate, and a quarter-wave plate. The reflected signal from the sample transmits through another quarter-wave plate, half-wave plate, and linear polarizer, and is then collected by a polarization-maintained single mode fiber and can be sent to detectors for further analysis. The spectra are recorded by a spectrometer (Princeton Instruments), consisting of a liquid nitrogen cooled CCD camera, with a monochromator grating (1714 g/mm). For the correlation and time-resolved measurements, two fiber-based superconducting single photon detectors (PhotonSpot) are used to collect the optical signals, which are then converted to electric signals and sent to the time-correlated single photon counter (Multiharp 300). We measure the resonance fluorescence second-order correlation in a standard Hanbury Brown and Twiss setup. The resonance fluorescence emission from the bound exciton is collected through the fiber and split by a 50:50 fiber beam splitter, and then sent to two single photon detectors. The analysis is assisted by Qucoa (PicoQuant) software.

*Cross-polarization setup*

In the resonant excitation experiments, we excite the emitter using linearly polarized light. The excitation beam from the tunable laser is sent through the polarization-maintained single-mode fiber and well-collimated by a spherical lens. The linear polarizer and the half-wave plate in the input path are used to align the input beam in a horizontal polarization state **H**. The reflection signals are collected in an orthogonal polarization state **V**, defined by the half-wave plate and the linear polarizer on the output optical path. Two quarter-wave plates are used to compensate the chromatic dispersion of the waveplate retardation. The optimized cross-polarization filter suppresses the input laser light with an extinction ratio exceeding $10^6$, measured by setting the filter in the cross-polarization state and the co-polarization state conditions.

## Funding declaration

The Waks group would like to acknowledge support from the AFOSR (grant #FA95502010250) and the Maryland-ARL quantum partnership (grant #W911NF1920181). The Pawlis group would like to acknowledge support from the Deutsche Forschungsgemeinschaft (DFG, German Research Foundation) under Germany's Excellence Strategy - Cluster of Excellence Matter and Light for Quantum Computing (ML4Q) EXC 2004/1−390534769. The funder played no role in study design, data collection, analysis and interpretation of data, or the writing of this manuscript.

## Acknowledgments

*Author contributions*

Y.J., R.P. and E.W. conceived the experiment. C.F., N.D. and Y.K. fabricated the device. Y.J. performed the experiment. C.F., A.A. and R.P. supported in setting up the experiment. Y.J. and J.B. analyzed the data. Y.J. and E.W. prepared the manuscript. All authors discussed the results and reviewed the manuscript. E.W. and A.P. supervised the experiment.

## Competing interests

The authors declare that they have no competing interests.

## Data and materials availability

All data needed to evaluate the conclusions in the paper are present in the paper and/or the Supplementary Materials.